# A FUZZY SIMILARITY BASED APPROACH FOR INTELLIGENT WEB BASED e-LEARNING

B.Senthilnayaki
Department of IT
UCEV, Anna University
Villupuram
nayaki_1@yahoo.com

K.Venkatalakshmi
Department of ECE
UCET, Anna University
Tindivanam
Venkata_krish@gmail.com

A.Kannan
Department of IST
CEG, Anna University
Chennai
akannan@annauniv.edu

## ABSTRACT
In this paper, an intelligent system for web based e-Learning is proposed which analyzes students' knowledge capacity by applying clustering technique. This system uses fuzzy logic and k-means clustering algorithm to arrange the documents according to the level of their performance. Moreover, a new domain ontology alignment technique is proposed that uses contextual information of the knowledge sources from the e-Learning domain for effective decision making. The proposed ontology alignment method has been empirically tested in an e-Learning environment and the experimental results show that the proposed method performs better than the existing methods in terms of precision and recall. The salient contributions of this paper are the use of Jaccard Similarity, fuzzy approach for ontology alignment and K-Means clustering algorithm for decision making using decision rules for providing intelligent e-Learning.

## 1. INTRODUCTION
E-Learning is a fast, which is easily accessible and non-linear learning process, which is now widely applied in distributed and dynamic environments such as the World Wide Web. Ontology plays an important role in capturing and disseminating the concept for effective human computer interactions. However, engineering of domain ontology is a labor intensive and time consuming process. In the past, some machine learning methods have been explored by various researchers for automatic or semi-automatic discovery of domain ontologies. Nevertheless, both the accuracy and the computational efficiency of these methods were not sufficient to make effective decision in e-learning applications.While constructing large scale ontology for real-world applications such as e-learning, the ability to monitor the progress of students' learning performance is a critical issue. Electronic learning (e-Learning) where students learn related course or other types of materials via online computer systems. It has proved to be an effective way of delivering materials to unreachable students.

One of the main advantages of e-Learning technology is that it can facilitate adaptive learning. The instructors can dynamically revise and deliver instructional materials in accordance with learners' current progress. The lecturer has to go through all the documents related to the subject and process them manually which an more time consuming process. In order to overcome this problem this work provides special features by building the ontology automatically. Using this system, to visualize similar group of students and to give the material to the student based on this result. So, this work aims at assisting the lecturers in a fast way by automatic construction of the ontology from a large document collection for effective decision making.

The e-Learning technologies can support automatic analysis of learners' and a methodology for automatically constructing concept maps to characterize learners' understanding for a particular topic has been designed and implemented. The concept map generation mechanism implemented in this work is underpinned by a context-sensitive text mining method and a fuzzy domain ontology extraction. Moreover, concept maps are useful to generate ontology and ontology provides an effective representation to represent concepts as well as the semantic relationships among concepts.

The main goal of this work is two folds. First, it involves on the development of a novel fuzzy domain ontology extraction method. Although some learning techniques have been proposed in the past for automatic or semiautomatic extraction of domain ontology the proposed work uses intelligent techniques to improve the accuracy in decision making. However all the existing methods are still primitive and further enhancement in terms of computational efficiency and learning accuracy is required.

In the e-learning scenario, concept has been recognized as the most important corporate asset and it is the key for understanding a particular topic and that has been designed and implemented to achieve sustainable competitive advantage [1]. An instructor can conduct adaptive teaching and learning based on the learners' knowledge structures as reflected on the concept maps, and can share this knowledge to improve the e-learning goal achievement. Domain ontology is one kind of feature which is used to represent the knowledge for a particular type of application domain. For example, knowledge about products, services, markets, etc. can be captured in an explicit and formal way such that it can be shared among human and computer systems.

Many researchers are already working on finding suitable techniques to capture and represent knowledge which can suit e-Learning [14,15]. Among then, G.A.Miller etal [2] proposed a data mining approach to discover the relations of the meta-data describing the various learning resources. In that work, the terms from the meta-data description files were parsed and stop words were removed. Moreover, language engineering tools such as WordNet were applied to extract the word roots Brill tagger algorithm was used for part of speech tagging.

Term Frequency(TF) and Inverse Document Frequency(IDF) heuristic developed in the field of Information Retrieval(IR) to extract prominent concepts from electronic messages generated in e-Learning [3,4]. In that system, knowledge density score was compified based on the TF/IDF term weighting formula to assess the extent of contribution to online knowledge sharing by individuals.

In [5], clustering was applied to find both groups of similar teams and similar individual members. In the case of mining, data collected from students using an online discussion board were used by them. Here, the classical k-means technique can be used to document clustering which requires relatively few clusters and to reduce the set of pre-selected words.

With this view in mind, the present work uses parts of speech tagging using wordnet. In this work, we use a document parsing approach which employs TF/IDF and other linguistic pattern recognition methods to extract concepts from text. In addition, we deal with the automatic construction of taxonomy of concepts as well. We use thresholds for measuring the understanding level of students. Moreover, we use Jaccard Similarity for measuring concept similarity and clustering technique to group similar documents into set up groups using k-mean clustering. Agents and rules are used in this work to analyze the students and to provide them suitable input based on the ontology built. Finally, we use Fuzzy Logic for classification while building the ontology.

This work proposes intelligent techniques software tool for enhancing e-Learning technology and shows how to apply the context-sensitive text mining method and the fuzzy domain ontology extraction algorithm to automatically generate concept maps to reveal the knowledge structures of students who are engaged in e-Learning. As a result, instructors can conduct adaptive teaching and learning based on the information disclosed on the ontology maps.

The major contributions of this work are the design of a new architecture with modules for pre-processing indexing, concept generation, similarity computation and grouping with for identifying relevant course contents. We use a knowledge base consisting of bright, medium and dull student details and the rules corresponding to these classifications. User interaction is provided to monitor the understanding level of the students so that the teacher can flexibility change his method of teaching.

## 2. PROPOSED WORK

The proposed System Architecture is shown in the figure 1and this architecture provides adaptable, interactive, distributed, collaborative, personalized and intelligent E-learning system. This system is used to support instructional design and to retrieve relevant information about the learners. It processes and analyzes the data results to enable meaningful e-learning recommendations. Most of the e-learning systems consider only one feature into account. So it is necessary to proposes a new e-learning System that supports all the features so that it can be considered as an efficient one.

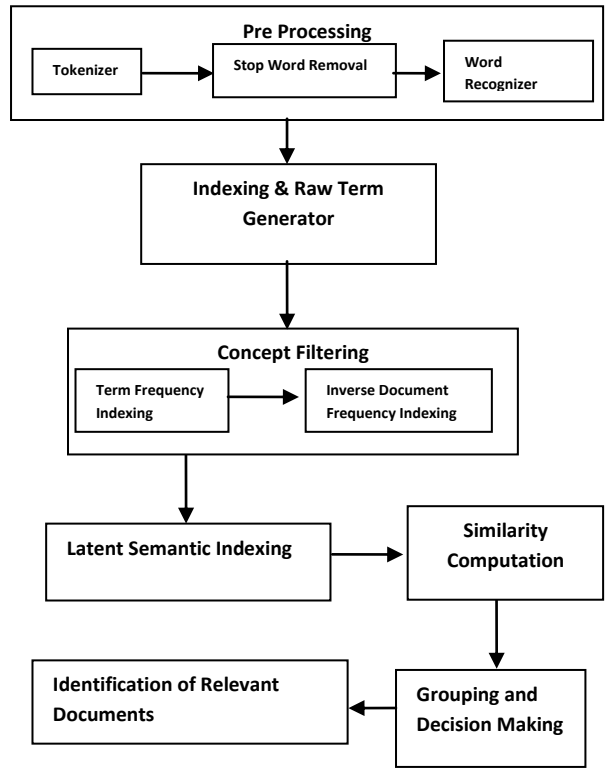

Figure.1 System Architecture

This work also contributes in similarity computation, clustering technique and the uses fuzzy logic for effective decision making in ontology construction.

## 3.1 Preprocessing of Learners Knowledge

Preprocessing basically consists of a process to optimize the list of terms that identify the collection. The pre-processing module is used to accept input text from the text corpora. The text file is converted into tokens using a tokeniser. Each of these tokens are passed to the stop word removal system where the stop words such determiners and prepositions determiners and prepositions are removed from the source documents. Since these words appear in any contexts and they cannot provide useful information to describe a domain concept they can be removed. In our implementation, a stop word file is constructed based on the standard stop word file. The words obtained after stop word removal are stored in another text file. The text file is read and is given to the POS tagging processes. In this process, identification of words as nouns, verbs, adjectives and adverbs using Wordnet is carried out so that they can be tagged with the part of speech.

## 3.2 Indexing and Raw Term Generation of Learners Knowledge

Indexing raw term generation gets the input from the pre-processing module as tokens. These tokens are divided into number of documents. Each document contains set of words (ie five to ten terms) and the occurrence of term in every document is to be computed. Finally, a matrix is generated to show the terms in rows and columns for the document. Finally an index is made to arrange the terms in ascending order.

## 3.3 Concept Filtering of Learners Knowledge

Concept filtering is performed TF and IDF. Using TF indexing it is possible to normalize the raw frequencies across a single document. For example, if a document had two words, one occurring twice and the other occurring thrice, the first word would be normalized to 2/5 (0.4) and the other to 3/5 (0.6). The term count in the given document is simply the number of times a given term appears in that document. This count is usually normalized to prevent a bias towards a document to give a measure of the importance of the term $t$ within the particular document $d$. Thus, we have the term frequency TF($t,d$).

Moreover, Inverse Document Frequency attempts to smooth out the frequency of a word across documents. If a word occurs in more than one document it means that it is less precise and hence its value should go down concept weight can be calculated by using the combination of TF and IDF as follows:

idf (t, D) = log |D| / | {d € D : t € d } |     ----- ( 1 )

where, D is the total number of document in this corpus, | {d € D: t € d} | is the number of documents where the term $t$ appears (i.e. tf(t,d) $\neq$ 0 ). If the term is not in the corpus, this will lead to a division-by-zero. It is therefore common to adjust the formula to  1+ | { d: t € d}|. Then tf*idf is calculated as:

tf  * idf (t,d,D) = tf (t,d) * idf (t,D)       ------ (2)

A high weight in tf*idf is reached by a high term frequency and a low document frequency of the term in the whole collection of documents; the weights hence tend to filter out common terms.

## 3.4 Latent Semantic Indexing

Latent Semantic Indexing(LSI) in [7] attempts to uncover latent relationships among documents based on word co-occurence. So if document A contains (w1,w2) and document B contains (w2,w3), we can conclude that there is something common between documents A and B. in this case w2 is common between A and B.

A = U * S * V$^T$           -------- (3)

A$_k$ = U$_k$ * S$_k$ * V$_k$$^T$           -------- (4)

Where, A is the original matrix, U is the word vector, S is the sigma vector, V is the document vector, U$_k$ is the reduced word submatrix consisting of 0..k-1 cols, S$_k$ is the reduced sigma submatrix consisting of 0..k-1 cols, 0..k-1 rows and V$_k$ is the reduced document submatrix consisting of 0..k-1 cols.

LSI uses Singular Value Decomposition (SVD) and gets three different matrices U, S and V. Once this is done, the three vectors are reduced and the original vector is rebuilt from the reduced vectors. Because of the reduction, noisy relationships are suppressed and relations become very clearly visible.

## 3.5 Similarity Computation

Similarity computation is carried out based on mutual information. Mutual Information is an information theoretic method to compute the dependency between two entities and is defined by [6]. Another method used for similarity measurement is Balanced Mutual Information (BMI) which is to compute the association weights among tokens [7].

In this paper, we use Jaccard computation [4] to compute the association weights among tokens. This method considers both term presence and term absence of the documents as the relationships. This is used to estimate the membership values of term belonging to a concept. Membership values are computed for the relation, attribute and concept are define in the fuzzy domain ontology. The formula of similarity measure using Jaccard's coefficient which is given by the following formula:

$$Jaccard's\ sim(A,B) = P(A \cap B)/P(A \cup B)$$
$$= \frac{P(A,B)}{P(A,B) + P(A,\overline{B}) + P(\overline{A},B)}$$

(5)

The Jaccard's coefficient is practically applied to define similarity measure in many systems which employs this formula for machine learning techniques to find mapping in the ontology model. This system maps the similarity of two ontologies into a new ontology [5]. A methodology of ontology mapping proposed in [3] is based on similarity measurement by Jaccard's coefficient as well.

The probabilities of the similarity measure of ontology structure in each document are computed using Jaccard's coefficient. The values of the probability will be between 0 and 1.

## 3.6 Similarity using Fuzzy Logic

Similarity measure based on Fuzzy Logic is proposed in [4] to measure for software project similarity. In this paper, we propose the methodology of classification of similarity of the learner's behaviour using Mamdani Fuzzy Model.

We use fuzzy rules to handle the uncertainty of the Learner learning Styles prediction. The learning styles are identified based on the learners' membership values for low/medium/high type of dimensions of learning style model.

**Table 1. Three Rules for Fuzzy**

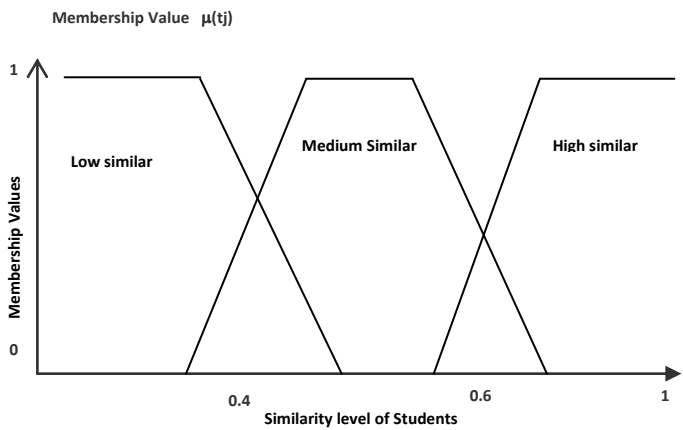

Figure 2. Similarity using Fuzzy Model Relation

In our design, three categories namely low, medium and high are use to classify fuzzy ontology. The fuzzy logic based similarity model proposed in this work is shown figure 2. We accept 0.5 as the threshold, both medium and higher similar level of similarity. For low similarity, the student is taught many more times with better learning materials and methods. For medium similarity, the learning material is explained to the students with more explanations. For higher similarity, the learning material is sent to the students' knowledge base for storage so that he can use it at all future times.

Table 1. Three Rules for Fuzzy

| Descriptions of Rules: |
| --- |
| If (similarity values is < 0.5) then ( low level students). |
| If (similarity values is ≤ 0.3 and > 0.7) then (medium level students). |
| If (similarity values is ≥ 0.6) then (high level students). |

The symmetric Mamdani fuzzy membership function used in the proposed model that identifies three categories of the learners (based on their learning styles) namely low level, medium level and high level. The proposed work has been evaluated using accuracy percentage considering the total numbers of sample documents. The table shown below provides the comparative values obtained from this work with respect to the other existing algorithms for 30, 60, 90, 120 and 150 numbers of learners. It is evident from the table that the proposed model shows an increased accuracy percentage compared to the other algorithm namely Bayesian algorithm. This is due to the fact that the fuzzy inference engine has been used in this work which is efficient in handling the uncertain information.

### 3.7 Grouping

This section explains how clustering how is used to convert very large number of documents into similar group of learner's behaviors. This approach uses k-means clustering algorithm to form efficient similar document groups. In this system, a new procedure is described that operates efficiently in high dimensions to producing clustering of similar documents.

Based on the grouping results, materials on the same subject are grouped as simple, medium and hard with respect to understanding. The decision model uses this grouping to provide suitable materials to the learners.

### 4.0 EXPERIMENTS AND RESULTS

Based on the clustering formed, materials are given to domain experts to develop ontologies for each group. When a learner learns a concept using this system, he/she is asked to develop an ontology or concept map. In case of new users, key words are obtained and ontology is created by the system. The ontology developed on the user's understanding is compared with the ontology generated by the domain expert and their Jaccard similarity measure is computed. If the similarity is above 75% the material from this group is delivered to the learner. This experiment is repeated with different groups for each user. Finally, the user is proposed with a suitable e-learning material.

In this work, small scale experiment of testing the functionality of the prototype system and the accuracy of the aforementioned domain ontology discovery method have been conducted under a e-Learning environment. A group of undergraduate in computer science and information technology students were recruited to try the prototype system. All of these learners have attended a course in knowledge management before. At the beginning of the experiment, they attended a briefing session of twenty minutes to learn the objective of this experiment. They were the instructed to write the most important concepts in knowledge management using concise and precise statements on an on-line discussion board. They were given fifteen minutes to write their answers. After this, the coordinator of the experiment executed the ontology discovery method to discover the domain concept. From the on-line answers which are about the writers' perception of concept management and using original ontology similarity computation is performed. Based on these experiments, the level of the students is judged and suitable course materials are provided to them.

In this experiment, we have tested the algorithms with our prototype, coding by in java and MySQL as a database. We have started with tokens which are divided into 8 words for each document. Then we computed the similarity of concepts, properties, and relations in each document based on similarity measurement by Jaccard's coefficient.

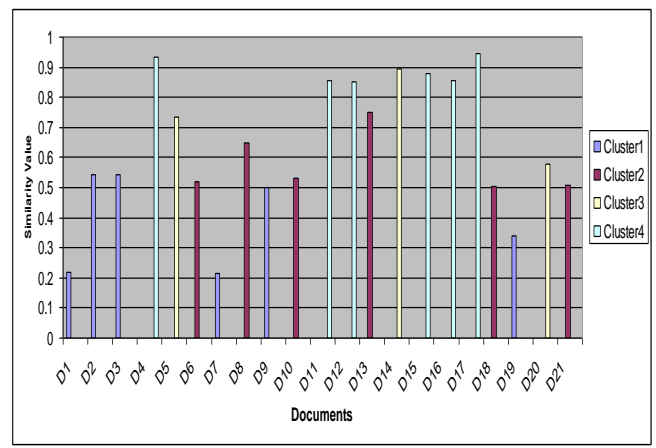

Figure 3. Bar Chart for Grouping Documents

The results obtained from the application of k-means clustering algorithm on the data derived from this e-learning environment is depicted in the form of an ontology in figure 3. From this figure, it can be observed that the four clusters formed namely c1, c2, c3 and c4 are the basis for the ontology developed.

Figure 4 shows the time analysis for retrieved documents of various size.

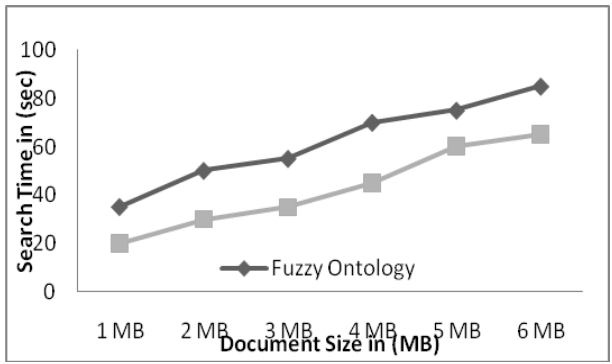

Figure 4 Time Analysis

From figure 4, it is observed that the search time is reduced when a fuzzy ontology is employed for data representation. This is due to the fact that fuzzy .

Figure 5 shows the bar chart representation of the relevancy comparisons between the existing ontology based approach and the proposed fuzzy ontology based approach.

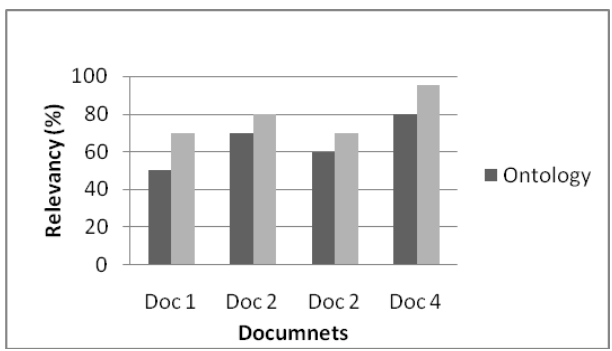

Figure 5 Relevancy Comparision

From figure 5, it can be seen that the relevancy in document retrieval for retrieving various documents are higher in the proposed fuzzy ontology based approach than the existing ontology based approach. This is due to the presence of fuzzy rules which compute the relevancy more accurate than the existing ontology based method.

Figure 6, shows the false positive analysis. From figure 6, it

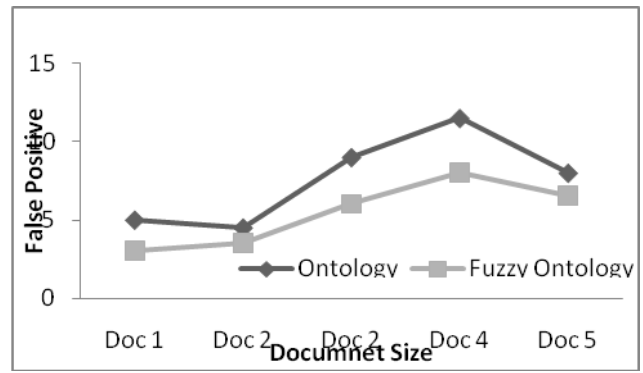

Figure 6 False Positive

can be observed the false positive rate is reduced in the proposed fuzzy ontology based approach when it is compared with the existing ontology based approach. This is due to the fact that uncertain data are handled efficiently by the fuzzy logic based approach when it is combined with ontology for effective decision making.

## 5.0 CONCLUSIONS AND FUTURE ENHANCEMENTS

In this paper, an intelligent system for concept map generation ontology construction and decision making in e-Learning is proposed. It has been designed and implemented using intelligent rules. It uses an online discussion board and classifies the learning materials presented to the student according to the fuzzy ontologies constructed. Moreover, the answer documents from the students are classified based on the ontology which computes similarity using fuzzy set, Jaccard's Similarity method and clustering techniques. This gives the more effective result to evaluate the quality of the document extracted. Using this system, instructors can quickly identify the learning status of their students. From this work, it is seen that fuzzy ontology provides a way to identify ontology similarities in documents which are very essential in e-Learning. Our preliminary experiment shows that the extracted fuzzy ontology similarity an accurately reflects the domain concept embedded in on-line discussion messages. Future work, involves larger scale of quantitative evaluation of our ontology extraction method with other similar methods.